\documentclass[twocolumn,superscriptaddress,aps,prb]{revtex4-2}

\usepackage{bm}
\usepackage{verbatim}
\usepackage{appendix}
\usepackage{graphicx}
\usepackage{amsmath}
\usepackage{amssymb}%
\usepackage{color}
\usepackage{ulem}
\usepackage{lipsum}
\usepackage{xurl}
\usepackage{multirow}
\usepackage[colorlinks=true,citecolor=blue,urlcolor=blue]{hyperref}
\usepackage{mathrsfs}
\usepackage{braket}
\usepackage{hhline}

\begin{document}

\title{Dissipation-driven topological phase transitions in open quantum systems independent of system Hamiltonian}
\author{Tian-Shu Deng}
\email{20220225@ynu.edu.cn}
\affiliation{School of Physics and Astronomy, Yunnan Key Laboratory for Quantum Information,
Yunnan University, Kunming 650091, China}
\author{Fan Yang}
\email{101013867@seu.edu.cn}
\affiliation{Key Laboratory of Quantum Materials and Devices of Ministry of Education, School of Physics, Southeast University, Nanjing 211189, China}
\begin{abstract}{
We investigate dissipation-driven topological phase transitions in one-dimensional quantum open systems governed by the Lindblad equation with linear dissipation operators, which ensure the density matrix retains its Gaussian form throughout the dynamics. By employing the modular Hamiltonian framework, we rigorously demonstrate that the $\rm{Z}_2$ topological invariant characterizing steady states in one-dimensional class D systems is exclusively dependent on the dissipation operators, rather than the system Hamiltonian. Through a sudden quench protocol where the system evolves from the steady state of one Lindbladian to another, we reveal that topological transitions can occur at analytically predictable critical times, even when the initial and final steady states share identical topological indices. These transitions are shown, both analytically and numerically, to depend solely on dissipation parameters. Entanglement spectrum analysis demonstrates bulk-edge correspondence in non-equilibrium density matrices via coexisting single-particle gap closures (periodic boundaries) and topologically protected zero modes (open boundaries), directly underpinning the detection of dissipation-induced topology in quantum simulators.
}
\end{abstract}

\maketitle

%\tableofcontents

\section{Introduction}

The investigation of topological insulators and superconductors in closed quantum systems constitutes a well-established research paradigm within condensed matter physics\cite{Topo1,Topo2,Topo3}. Here, the topological properties of a Hamiltonian are intrinsically encoded in its eigenstates, exemplified by bulk-boundary correspondence and quantized invariants like Chern numbers derived from Berry phases. However, the situation becomes fundamentally different in open quantum systems governed by Lindblad dynamics\cite{Lindblad1,Lindblad2}. Here, the system's evolution is described by a Liouvillian superoperator rather than a Hamiltonian, and quantum states must be treated as mixed density matrices. Two distinct paradigms emerge for characterizing topology in such dissipative settings. One focuses on the intrinsic topology of mixed states through their density matrix structure\cite{MixTopo1,MixTopo2,MixTopo3,MixTopo4,MixTopo5,MixTopo6,MixTopo7,MixTopo8,Huang25}, while the other explores the non-Hermitian topology\cite{NHTopo1,NHTopo2,NHTopo3,Kawabata19,NHTopo5,Lieu20,NHTopo6,NHTopo7} encoded in the Liouvillian superoperators. In parallel, field-theoretic approaches have established a comprehensive classification and the bulk-boundary correspondence for Gaussian steady states across all symmetry classes and dimensions\cite{steadystateclass1,Huang25}, with extensions to interacting systems\cite{steadystateinter}.
Recent theoretical advances have established the modular Hamiltonian formalism ($\hat{\rho} = e^{-\hat{G}}$) as a rigorous framework for characterizing topology in quadratic Lindbladian systems\cite{Altland21,ModTopo1,ModTopo2,ModTopo3}, where $\hat{\rho}$ remains Gaussian and $\hat{G}$ stays quadratic. The Hermitian nature of $\hat{G}$ allows direct application of topological classification schemes of closed systems, provided its symmetry class remains preserved during dissipative evolution. This preservation imposes strict requirements for both Lindbladian operators $\{\hat{L}_\mu\}$ and the system Hamiltonian. There exists symmetry classification correspondence between Gaussian states and quadratic Lindbladians \cite{Mao25}. While existing studies have established that topological transitions necessarily occur at finite critical times $t_0$ when initial and steady states occupy distinct topological sectors, it remains unknown whether multiple topological transitions can occur even when initial and final states are in different topological sectors, or how to determine their exact occurrence times in symmetry-constrained open systems.

Here, we study one-dimensional Lindblad systems with Majorana fermions, where the Hamiltonian takes a quadratic form and dissipation operators are linear in Majorana operators. The modular Hamiltonian of such systems inherently satisfies particle-hole symmetry, naturally placing it within the class D framework without time-reversal symmetry\cite{Topo3}. Crucially, we find the steady state depends exclusively on the dissipation operators rather than the system Hamiltonian. 

To probe topological transitions, we employ a sudden quench protocol. The system is prepared in the steady state of an initial Lindbladian $\mathcal{L}_i$ and is then evolved under a constant post-quench Lindbladian $\mathcal{L}_f$.
We demonstrate that even when initial and final states share the same $Z_2$ topological indices, the dissipative dynamics can drive multiple topological transitions at precisely predictable time points. These critical times are fully determined by the dissipation operators, revealing a complete decoupling from Hamiltonian-driven dynamics observed in closed systems.

The entanglement spectrum (ES), constructed from the reduced density matrix, serves as a critical probe for detecting phase transitions in dissipative dynamics\cite{ESLind1,ESLind2,ESLind3} and quench dynamics\cite{ESQuench1,ESQuench2,ESQuench3,ESQuench4,ESQuench5}. For Gaussian states, the full ES can be directly reconstructed from the single-particle entanglement spectrum (SPES), which is derived from correlation matrices\cite{SPES1,SPES2}. In closed quantum systems, SPES gap-crossing behavior serves as a hallmark of stable topological structures in quench dynamics \cite{ESQuench3,ESQuench4}. Extensions to open systems further connect SPES gap transitions with non-Hermitian topology in Lindblad frameworks. Our work demonstrates that dynamical topological phase transitions in modular Hamiltonians exhibit two universal signatures. The SPES gap universally closes at critical points under periodic boundary conditions throughout dynamical evolution, while open boundaries sustain topologically protected zero modes that mirror the modular Hamiltonian’s edge spectrum. These findings establish a bulk-edge correspondence in modular dynamics and provide experimentally measurable criteria for identifying its topology. By directly linking modular Hamiltonian features to entanglement observables, our results enable dynamical detection of entanglement-driven topology in quantum simulation platforms.

The rest of the paper is organized as follows. Sec.~\ref{sec:model} establishes a 1D quadratic Lindblad model for spin-1/2 fermions using Majorana representation, defines the correlation matrix linked to Gaussian states' modular Hamiltonian, and characterizes topology through a $Z_2$ Pfaffian invariant. Sec.~\ref{sec:steadystate} derives analytical expressions for steady-state topological numbers in open quantum systems, demonstrating their exclusive dependence on dissipation operators. Sec.~\ref{sec:Topophasedyn} identifies explicit temporal transition points for dissipation-induced topological phase transitions during dynamics. Sec.~\ref{sec:Entanglement} analyzes entanglement spectrum evolution under time evolution. Sec.~\ref{sec:sum} concludes with key findings.

\section{Model}
\label{sec:model}
We start from a quadratic Lindbladian in one-dimensional lattice with spin-1/2 fermions, written as
\begin{eqnarray}
\frac{d\hat{\rho}}{dt}=-i[\hat{H},\hat{\rho}]+\sum_{\mu=1}^{N}\left(2\hat{L}_{\mu}\hat{\rho}\hat{L}_{\mu}^{\dagger}-\left\{ \hat{L}_{\mu}^{\dagger}\hat{L}_{\mu},\hat{\rho}\right\} \right).
\label{eqlind}
\end{eqnarray}
Here, the quadratic Hamiltonian in Majorana reprensentation is given by
\begin{equation}
\label{Hkit}
\hat{H}=\sum_{m,n=1}^N\sum_{s,s'=1,2} K_{m,s,n,s'} \hat{w}_{m,s} \hat{w}_{n,s'},
\end{equation}
where $K$ is imaginary and antisymmetric matrix. We have defined the Majorana operators as $\hat{w}_{m, 1}=\frac{1}{\sqrt{2}}\left(\hat{c}_{m}^{\dagger}+\hat{c}_{m}\right)$ and $\hat{w}_{m, 2}=i \frac{1}{\sqrt{2}}\left(\hat{c}_{m}-\hat{c}_{m}^{\dagger}\right)$, and commutation relation is written as $\{\hat{w}_{m,s},\hat{w}_{n,s'}\}=\delta_{m,n}\delta_{s,s'}$.
The linear Lindblad operator $\hat{L}_\mu$ in Eq.(\ref{eqlind}) is superposition of both annihilation operators and creation operators, which takes the form
\begin{equation}
\label{Ln}
\hat{L}_{\mu}=\sum_{m=1,s}^{N}C_{\mu,m,s}\hat{w}_{m,s}.
\end{equation}

A distinctive feature of the quadratic Lindblad equation is that if the initial state is a Gaussian state, the density matrix will remain the Gaussian form ,
\begin{equation}
\label{rhoGZ}
\hat{\rho}=e^{-\hat{G}}=e^{-\frac{i}{4}\sum_{m,s,n,s'}G_{m,s,n,s'}{\hat{w}}_{m,s}{\hat{w}}_{n,s'}}/Z,
\end{equation}
where $Z$ is the normalization constant which ensures $\rm{Tr}({\hat{\rho}})=1$. The matrix $G$ is real and anti-symmetric and the Hermitian operator $\hat{G}$ is known as the modular Hamiltonian of density operator. The topological characteristics of the modular Hamiltonian provide a framework for identifying the topology of the associated density operator, which is related to the single-particle correlation matrix\cite{quaninfgeo} as
\begin{equation}
\label{delG}
\begin{aligned}
\Delta=\tanh{(i\frac{G}{4})},
\end{aligned}
\end{equation}
and the single-particle correlation matrix is given by \begin{equation}
\label{delijs}
\begin{aligned}
\Delta_{m,s,n,s'}={\rm Tr}\{\hat{\rho}(t)[\hat{w}_{m,s},\hat{w}_{n,s'}]\}. 
\end{aligned}
\end{equation}
As governed by Eq.~\eqref{delG}, the matrices \(\Delta\) and \(G\) are topologically equivalent. Consequently, we characterize the topology of the density operator through the topological invariants encoded in the \(\Delta\) matrix.

By substituting Eq.~(\ref{delijs}) into Eq.~(\ref{eqlind}), we obtain the time-evolution equation of the correlation matrix $\Delta$ \cite{cmmajdt1,cmmajdt2,cmmajdt3}as
\begin{equation}
\label{delt}
\begin{aligned}
\frac{d\Delta(t)}{dt}=-i[H_{{\rm eff}}\Delta-\Delta H_{{\rm eff}}^{\dagger}]-4Y,
\end{aligned}
\end{equation}
where we have defined $H_{{\rm eff}}=2(K-iX)$, $M_{p,q}=\sum_{\mu}C_{\mu,p}C_{\mu,q}^{*}$, $X=\frac{M+M^{T}}{2}$, and $Y=\frac{M-M^{T}}{2}$. Note that the matrices $\Delta$, $Y$, and $K$ are all anti-symmetric and the matrix $X$ is symmetric according to their definitions. Considering the lattice translational symmetry, we transform the correlation matrix $\Delta$ into quasimomentum space by defining $\hat{w}_{k,s}=\sum_{m}\hat{w}_{m,s}e^{ikm}$ and
\begin{equation}
\Delta_{s,s'}(k)=\sum_{m,n}\Delta_{m,s,n,s'}e^{ikm}e^{-ikn}={\rm Tr}\{\hat{\rho}(t)[\hat{w}_{k,s},\hat{w}_{-k,s'}]\}.
\end{equation} 
Then the evolution of correlation matrix is rewritten as
\begin{equation}
\label{delt}
\begin{aligned}
\frac{d\Delta(k)}{dt}=-i[H_{{\rm eff}}(k)\Delta(k)-\Delta(k)H_{{\rm eff}}^{\dagger}(k)]-4Y(k),
\end{aligned}
\end{equation}
where 
\begin{equation}
\label{Heffkx}
H_{\rm{eff}}(k)=2[K(k)-iX(k)].
\end{equation}
The matrices $K$, $X$ and $Y$ are all transformed into the quasimomentum space as $K_{s,s'}(k)=\sum_{m,n}K_{m,s,n,s'}e^{ikm}e^{-ikn}$, $X_{s,s'}(k)=\sum_{m,n}X_{m,s,n,s'}e^{ikm}e^{-ikn}$, $Y_{s,s'}(k)=\sum_{m,n}Y_{m,s,n,s'}e^{ikm}e^{-ikn}$. The anti-symmetry of the matrices $K$, $Y$ and $\Delta$ and the symmetry of the matrix $X$ acts on  quasimomentum space as $K(k)=-K^T(-k)$, $Y(k)=-Y^T(-k)$, $\Delta(k)=-\Delta^T(-k)$ and $X(k)=X^T(-k)$.

Since $\Delta$ is anti-symmetric, the modular Hamiltonian possesses an intrinsic particle-hole symmetry. In our context, we focus on D class which exhibits particle-hole symmetry but lacks chiral symmetry. The topological properties of one-dimensional class D systems can be characterized by a $\rm{Z}_2$ topological index $\nu$, which is calculated by Pfaffians of $\Delta$ \cite{M0Mp} as
\begin{equation}
\nu=M_0M_\pi,
\label{nuph}
\end{equation}
with 
\begin{equation}
M_{0/\pi}={\rm{sgn}}[-i{\rm{Pf}}\Delta(0/\pi)].
\label{M0p}
\end{equation}
When $M_0$ and $M_\pi$ have opposite signs, the topological invariant evaluates to $\nu = -1$, indicating a topologically non-trivial phase. Conversely, when $M_0$ and $M_\pi$ share the same sign, the invariant becomes $\nu = 1$, corresponding to a topologically trivial phase.
Given that the topological invariant $\nu$  only depends on $\Delta(k)$ of $k=0,\pi$, we introduce the notation $k_s\in\{0,\pi\}$ to denote these high-symmetry momenta points. 
The correlation matrix $\Delta(k)$ is Hermitian, and at the high-symmetry momenta points $\Delta(k_s)$ is anti-symmetric. These constrains  ensures the $2\times2$ matrix $\Delta(k_s)$ is propotional to $\sigma_y$, and the general form is given by
\begin{equation}  
\label{delPf}
\Delta(k_s) = i{\rm{Pf}}[\Delta(k_s)]\sigma_y. 
\end{equation}  
Here, ${\rm{Pf}}[\Delta(k_s)] \in \mathbb{R}$ is real-valued ,which decide the $Z_2$ topological invariant of modular Hamiltonian.

In the subsequent discussion, we consider a scenario where a steady state of one Lindbladian serves as the initial state. We investigate its dynamical evolution under a different Lindbladian and analyze topological phase transitions during this process. Dynamical topological phase transitions in dissipative systems are intimately connected to the topological index of the initial state and the steady states of the Lindbladians, which will be addressed in the next section.

\section{Topological phase diagram of steady states}
\label{sec:steadystate}

The steady-state solution for $\Delta(k)$ is obtained by setting $\frac{d\Delta}{dt} = 0$ in Eq.~\eqref{delt}, which yields 
\begin{equation}  
\label{delsk}  
H_{\text{eff}} \Delta_s - \Delta_s H_{\text{eff}}^\dagger = 4iY.  
\end{equation}  
Through vectorizing matrices $\Delta$ and $Y$ and expressing $H_{\text{eff}}$ in its eigenbasis $\{|\psi_\alpha\rangle\}$ with eigenvalues , the steady-state correlation matrix of this equation transforms into 
%Through vectorization with $|\Delta\rangle = \sum_{i,j} \Delta_{ij} |i\rangle \otimes |j\rangle$ and $|Y\rangle = \sum_{i,j} Y_{ij} |i\rangle \otimes |j\rangle$, this equation transforms into  
%\begin{equation}  
%|\Delta\rangle = 4i \left[ H_{\text{eff}} \otimes I - I \otimes H_{\text{eff}}^* \right]^{-1} |Y\rangle.  
%\end{equation}  

\begin{equation}  
\label{Delsksum}
\Delta_s(k) = 4i \sum_{\alpha,\beta=\pm} \frac{\langle \chi_\alpha | Y(k) | \chi_\beta \rangle}{\lambda_\alpha - \lambda_\beta^*} |\psi_\alpha\rangle \langle \psi_\beta|,  
\end{equation}  
where $|\psi_\alpha\rangle$ and $|\chi_\alpha\rangle$ denote the right and left eigenvectors of $H_{\text{eff}}$ respectively and $\lambda_\alpha$ ($\alpha = \pm$) denotes the eigenvalues of  $H_{\rm{eff}}$. Detailed deduction of Eq.~\eqref{Delsksum} is shown in Appendix \ref{app:Delsksum}. Note that the derivation of Eq.\eqref{Delsksum} requires the prerequisite that $\lambda_\alpha - \lambda_\beta^* \neq 0$ for all eigenvalues, a condition which is equivalent to a non-vanishing dissipative gap\cite{cmmajdt2}. This prerequisite is assumed to hold throughout our work.

At high-symmetric points, we have $H_{\text{eff}}(k_s) = 2[K(k_s) - iX(k_s)]$, where $K(k_s)$ is an antisymmetric Hermitian matrix and $X(k_s)$ is a symmetric Hermitian matrix. Therefore, $H_{\text{eff}}(k_s)$ could be written as
\begin{equation}  
\label{Heffhx}
H_{\rm{eff}}(k_s) = h_y(k_s) \sigma_y + ih_0(k_s)  I +i h_x(k_s)  \sigma_x + i h_z(k_s)  \sigma_z,  
\end{equation}  
where we have defined 
\begin{align}
&h_0(k_s)=-{\rm{Tr}}[X(k_s)],\quad h_x(k_s)=-{\rm{Tr}}[X(k_s)\sigma_x],\nonumber\\
&h_y(k_s)={\rm{Tr}}[K(k_s)\sigma_y],\quad h_z(k_s)=-{\rm{Tr}}[X(k_s)\sigma_z],
\end{align}
and $h_0(k_s),h_x(k_s), h_y(k_s), h_z(k_s)$ are all real-valued. The eigenvalues of $H_{\rm{eff}}(k_s)$ is written as
\begin{align}  
&\lambda_\pm(k_s) = ih_0(k_s) \pm E(k_s), \nonumber\\
&E(k_s) = \sqrt{h^2_y(k_s) - h^2_x(k_s) - h^2_z(k_s)},  
\end{align}
where $E(k_s)$ is purely real for $h^2_y(k_s) - h^2_x(k_s) - h^2_z(k_s)>0$ and purely imaginary for $h^2_y(k_s) - h^2_x(k_s) - h^2_z(k_s)<0$. The region where $E(k_s)$ is purely real(imaginary) is generally termed the region with preserved(broken) parity-time(PT) symmetry \cite{PT1,PT2}.
%Given Eq.\eqref{Heffhx}, it follows directly that $H^*_{\rm{eff}}(k_s) = -H_{\rm{eff}}(k_s)$.

Since $Y(k_s)$ is also Hermitian and anti-symmstric, it yields $Y(k_s)=y_{k_s}\sigma_y$ with $y_{k_s}={\rm{Tr}}[Y(k_s)\sigma_y]/2$. Then the Pfaffian number of correlation matrix $\Delta(k_s)$ is derived as  
\begin{equation}  
\label{Mkchipsi}
{\rm{Pf}}[i\Delta_s(k_s)]=\frac{1}{2}{\rm{Tr}}[\Delta(k_s)\sigma_y]=  \sum_{\alpha,\beta}  \frac{2iy_{k_s}f_{\alpha,\beta}}{\lambda_\alpha - \lambda_\beta^*},  
\end{equation}  
where we have defined
\begin{equation}  
f_{\alpha,\beta}=\langle \chi_\alpha | \sigma_y | \chi_\beta \rangle \langle \psi_\beta | \sigma_y | \psi_\alpha \rangle.
\label{fmnchipsi}
\end{equation}  
Given Eq.~\eqref{Heffhx}, it follows directly that $H^*_{\rm{eff}}(k_s) = -H_{\rm{eff}}(k_s)$. Thus we find $f_{+,+}=f_{-,-}=1(0)$ and $f_{+,-}=f_{-,+}=0(1)$ for the PT-symmetry-preserved (broken) regime. We then substitute the respective overlap factors $f_{\pm,\pm}$ of both regimes into the Pfaffian's expression Eq. \eqref{Mkchipsi}. 
Since $\lambda_{\alpha}(k_s)-\lambda_{\beta}^*(k_s)=2ih_0$ when $\alpha=\beta$ ($\alpha\neq\beta$) holds for the PT-symmetry-preserved (broken) regime
this substitution process yields identical expressions for the Pfaffian of the correlation matrix in both regimes as
\begin{equation}  
\label{delt0pff}
\mathrm{Pf}[i\Delta_s(k_s)] =  \frac{2y_{k_s}}{h_0(k_s)}.
\end{equation}
This result shows that the steady-state topology in one-dimensional class D is entirely controlled by the dissipation term rather than system Hamiltonian. The detailed proof of Eq.~\eqref{delt0pff} is given in Appendix \ref{app:delt0pff}.

The denominator $h_0(k)$ in Eq.\eqref{delt0pff} is explicitly given by  
\begin{equation}  
h_{0}(k) = -\text{Tr}[X(k)] =-\sum_{s} X_{s,s}(k).  
\end{equation}  
Combining the Fourier transformation of $X(k)$ with the definitions of $X$ and $M$, we derive  
\begin{equation}  
\label{h0}  
h_{0}(k) = -\frac{1}{2}\sum_{\mu, m, n, s} \left[ C_{\mu,m,s} C_{\mu,n,s}^* + C_{\mu,n,s} C_{\mu,m,s}^* \right] e^{ikm} e^{-ikn}.  
\end{equation}  
Introducing the momentum-space coefficients  
\begin{equation}  
\tilde{C}_{\mu,s}(k) = \sum_{m} C_{\mu,m,s} e^{ikm},  
\end{equation}  
the expression for $h_0(k)$ is simplified to  
\begin{equation}  
\label{h02}  
h_{0} = -\frac{1}{2} \sum_{\mu,s} \left[ \tilde{C}_{\mu,s}(k) \tilde{C}_{\mu,s}^*(k) + \tilde{C}_{\mu,s}(-k) \tilde{C}_{\mu,s}^*(-k) \right].  
\end{equation}  
which reveals the quantity $h_0$ is negative real values and a straightforward result is obtained as $M_{0/\pi}={\rm{sgn}}(y_{0/\pi})$.  

In conclusion, for both PT-symmetric regime and the PT-broken regime, the topological invariant reduces to  
\begin{equation}  
\nu =M_0M_\pi={\rm{sgn}}[y(0)]{\rm{sgn}}[y(\pi)],  
\end{equation}  
demonstrating the universal independence of ${\rm{Z}}_2$ topological invariant on system Hamiltonian $\hat{H}$.

%Figure \ref{fig:fig1} presents the complete topological classification of steady states using the ordered pair $(M_0,M_\pi)$. The system exhibits two distinct phases: (i) A trivial phase with $\nu=+1$ corresponding to $(M_0,M_\pi)=(\pm1,\pm1)$, where both symmetry indicators share the same sign; (ii) A topologically non-trivial phase with $\nu=-1$ characterized by $(M_0,M_\pi)=(\pm1,\mp1)$, where the symmetry indicators exhibit opposite signs. This dichotomous classification reveals the emergence of non-equilibrium topological order in the dissipative steady state.

\begin{figure}[t]
\includegraphics[width=8cm]{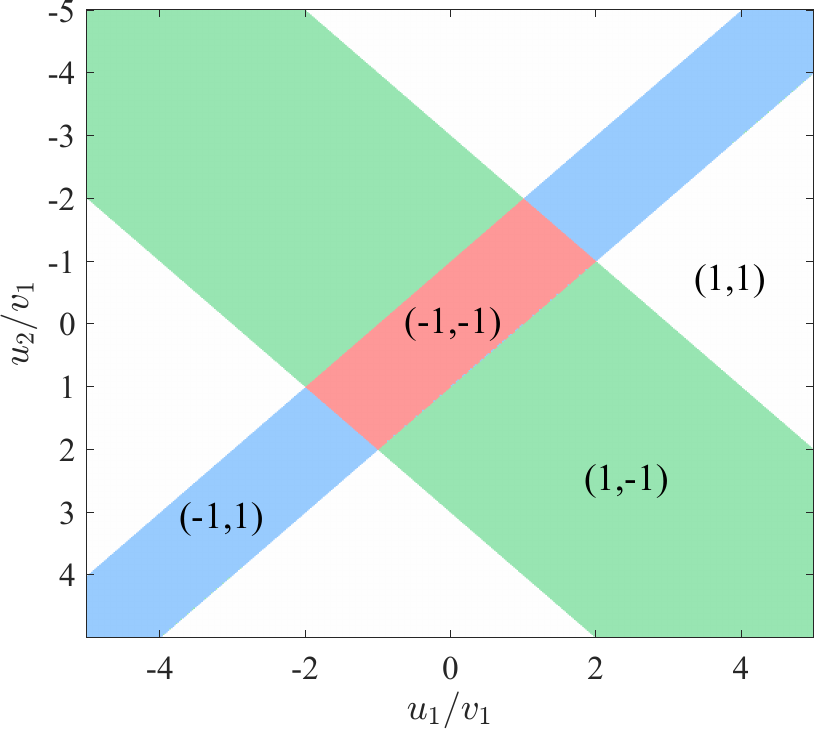}
\caption{Topological phase diagram of Lindblad steady state. Different colors denote different $(M_0,M_\pi)$. Red area  represents $(M_0,M_\pi)=(-1,-1)$, blue area  represents $(M_0,M_\pi)=(-1,1)$, green area represents $(M_0,M_\pi)=(1,-1)$ and white area represents $(M_0,M_\pi)=(1,1)$. Also, we have taken $v_2/v_1=-2$, and the dimensionless parameter $\gamma=0.1$. Note that the system Hamiltonian parameters have no influence on the topological properties of the steady state, although the validity of this phase diagram requires that the Hamiltonian is chosen to ensure a non-vanishing dissipative gap.}
\label{fig:fig1}
\end{figure}

Although the topological index $\nu$ only relies on $y(k=0/\pi)$, the phase diagram of the Lindblad steady state still has rich structures. For example, if we take the specific form of Lindblad operator as
\begin{align}
\hat{L}_{\mu}=\sum_{n}\sqrt{\gamma}[(u_{1}\delta_{\mu,n}+u_{2}\delta_{\mu+1,n})\hat{c}_{n}+(v_{1}\delta_{\mu,n}+v_{2}\delta_{\mu+1,n})\hat{c}_{n}^{\dagger}],
\end{align}
then $y(k)$ is given by
\begin{align}
y(k)&=\gamma[|u_{1}|^{2}+|u_{2}|^{2}+u_{2}u_{1}^{*}\cos k+u_{1}u_{2}^{*}\cos k\nonumber\\
&-(|v_{1}|^{2}+|v_{2}|^{2}+v_{2}v_{1}^{*}\cos k+v_{1}v_{2}^{*}\cos k)],
\end{align}
and the topological index takes the form as
\begin{align}
\nu&=M_{0}M_{\pi}\nonumber\\
&={\rm sgn}\{[(u_{1}+u_{2})^{2}-(v_{1}+v_{2})^{2}][(u_{1}-u_{2})^{2}-(v_{1}-v_{2})^{2}]\}.
\end{align}
In Fig. \ref{fig:fig1}, we present the phase diagram of the dissipative steady state, constructed by fixing $v_2/v_1$ and varying $u_2/v_1$ and $u_1/v_1$. Here, distinct topological phases are  classified based on the Pfaffian signature $(M_0,M_\pi)$. The phase characterized by $\nu=1$, where Pfaffians of $k=0$ and $k=\pi$ exihibit identical signs, represents the topologically trivial phase. Conversely, the phase with $\nu=-1$, where Pfaffians of $k=0$ and $k=\pi$ exihibit opposite signs, signifies the topologically non-trivial phase. 

\section{Topological phase transition in the dissipative dynamics}
\label{sec:Topophasedyn}

We investigate the temporal evolution of the modular Hamiltonian's topological number in dissipative quantum dynamics. The density matrix is initialized as the steady state of the initial Lindbladian $\mathcal{L}_i$, whose correlation matrix is given by $\Delta_{s,i}(k_s)$. Thereafter, the initial state evolves under the post-quench Lindbladian $\mathcal{L}_f$ towards the final steady state, whose correlation matrix is given by $\Delta_{s,f}(k_s)$. In this protocal, the time-dependent correlation matrix $\Delta(k_s,t)$ at the high-symmetry points evolves as
\begin{align}
\label{deltakt}
&\Delta(k_s,t)-\Delta_{s,f}(k_s)=\sum_{\alpha,\beta}\exp[-i(\lambda_{\alpha,f}-\lambda_{\beta,f}^{*})t]\nonumber\\
&\times\langle\chi_{\alpha,f}|[\Delta_{s,i}(k_s)-\Delta_{s,f}(k_s)]|\chi_{\beta,f}\rangle|\psi_{\alpha,f}\rangle\langle\psi_{\beta,f}|,
\end{align}
where $\lambda_{\alpha,f}$ is the eigenvalue of the effective non-Hermitian Hamiltonian in the post-quench Lindbladian $\mathcal{L}_f$, with $|\psi_{\alpha,f}\rangle$ and $|\chi_{\alpha,f}\rangle$ being its respective right and left eigenvectors.  According to Eq.~\eqref{nuph}, the topological invariant of the evolving density state is directly calculated from the Pfaffians of correlation matrix 
\begin{align}
\label{PfiDelkt}
{\rm{Pf}}[i\Delta(k_s,t)]=\frac{1}{2}{\rm{Tr}}[{\sigma_y\Delta(k_s,t)}].
\end{align}
Combining Eq.~\eqref{deltakt}, Eq.~\eqref{PfiDelkt} and Eq.~\eqref{fmnchipsi}, the Pfaffians is derived as
\begin{align}
\label{deltakPf}
{\rm Pf}[i\Delta(k_{s},t)]=&{\rm Pf}[i\Delta_{s,f}(k_{s})]+\frac{1}{2}\sum_{\alpha,\beta}\exp[-i(\lambda_{\alpha,f}-\lambda_{\beta,f}^{*})t]\nonumber\\
&\times{\rm \{{Pf}}[i\Delta_{s,i}(k_{s})]-{\rm {Pf}}[i\Delta_{s,f}(k_{s})]\}f_{\alpha,\beta}.
\end{align}
For the PT-symmetry-preserved regime, we only need to consider the $\alpha=\beta$ case which yields $\lambda_\alpha-\lambda_\beta^*=2ih_{0,f}$. And for the PT-symmetry-broken regime, we only consider the $\alpha \neq \beta$ case also yields $\lambda_\alpha-\lambda_\beta^*=2ih_{0,f}$. In both scenarios, Eq.~\eqref{deltakPf} is simplified to
\begin{align}
\label{Psigamma}
{\rm{Pf}}[i\Delta(k_s,t)]=\exp(2h_{0,f}t)\left(\frac{2y_{k_s,i}}{h_{0,i}}-\frac{2y_{k_s,f}}{h_{0,f}}\right)+\frac{2y_{k_s,f}}{h_{0,f}}.
\end{align}
A dynamical topological phase transition takes place at points when topological index $\nu(t)=M_0M_\pi={\rm{sgn}}\{{{\rm{Pf}}[i\Delta(0,t)]{\rm{Pf}}[i\Delta(\pi,t)]}\}$ flips, which means either ${\rm{Pf}}[i\Delta(0,t)]$ or ${\rm{Pf}}[i\Delta(\pi,t)]$ crosses zero. The equation ${\rm{Pf}}[i\Delta(k_s,t)]=0$ has at most one real solution, given by
\begin{align}
\label{tpif}
t_{p,k_s}=\frac{1}{2h_{0,f}}\log{\frac{\frac{y_{k_s,f}}{h_{0,f}}}{\frac{y_{k_s,f}}{h_{0,f}}-\frac{y_{k_s,i}}{h_{0,i}}}}.
\end{align}
The real positive solution of transition time $t_{p,k_s}$ exists when $\frac{\frac{y_{k_s,f}}{h_{0,f}}}{\frac{y_{k_s,f}}{h_{0,f}}-\frac{y_{k_s,i}}{h_{0,i}}}>1$, which necessarily requires the dissipation coefficients $y_{k_s,i}$ and $y_{k_s,f}$ to have opposite signs. 
%The topological index $\nu(t)=M_0M_\pi={\rm{Pf}}[\Delta(0,t)]{\rm{Pf}}[\Delta(\pi,t)]$ undergoes discrete jumps when either ${\rm{Pf}}[\Delta(0,t)]$ or ${\rm{Pf}}[\Delta(\pi,t)]$ crosses zero. 
The multiplicity of phase transitions depends on the topological index configuration $(M_0,M_\pi)$ of the pre-quench and post-quench Lindbladian steady states. No phase transitions occur when neither $M_0$ ($M_{0,i}\rightarrow M_{0,f}$) nor $M_\pi$ ($M_{\pi,i}\rightarrow M_{\pi,f}$) undergoes a sign reversal. A single transition at $t_{p,0/\pi}$ arises if $M_{0/\pi}$ reverses sign while the sign of $M_{\pi/0}$ remains unchanged. When both $M_{0}$ and $M_{\pi}$ undergo sign reversals, two topological phase transitions (at $t_{p,0}$ and $t_{p,\pi}$) emerge, where the initial state and the final state share the same topological invariant $\nu=M_{0}M_{\pi}$.

%two transitions occur when both $M_{0,i}\neq M_{0,f}$ and $M_{\pi,i}\neq M_{\pi,f}$, a single transition appears for partial mismatch, while no transitions occur when the initial and final topological indices coincide completely.

\begin{figure}[t]
\includegraphics[width=8cm]{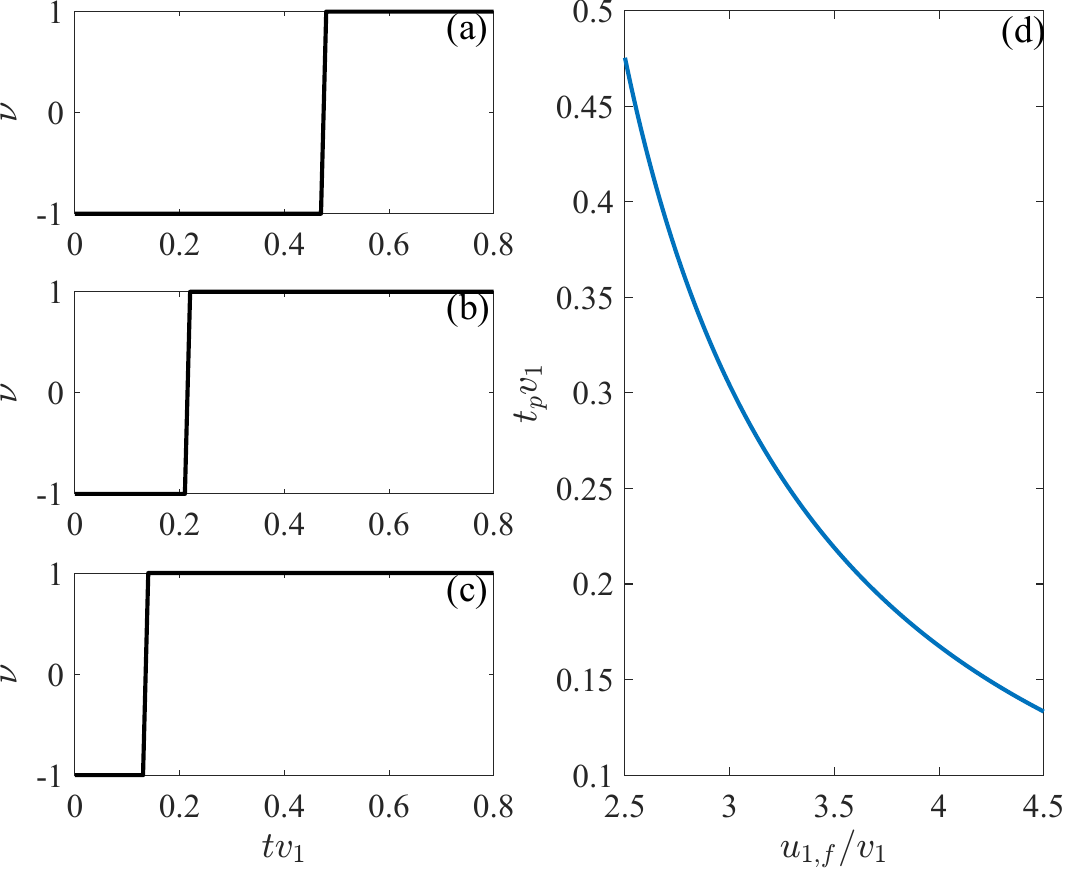}
\caption{(a)(b)(c)Topological phase transition characterized by the modular Hamiltonian's topological invariant $\nu(t)$ following a dissipative quench from Lindbladian steady state of  $(M_0,M_\pi)=(1,-1)$ topological phase to those of $(M_0,M_\pi)=(1,1)$ phase. 
During the entire quench process, $v_2$ and $v_1$ remain constant, the ratio $v_2/v_1$ is maintained at $-2$, and the dimensionless parameter $\gamma$ takes the value $0.1$.
Initial density matrix is the steady state of the Lindbladian with parameters $u_{1,i}/v_1=3$, $u_{2,i}/v_1=3$ and the post-quench Lindbladian parameters are taken as $u_{1,f}/v_1=2.5, u_{2,f}/v_1=-1$ for (a), $u_{1,f}/v_1=3.5, u_{2,f}/v_1=-1$ for (b), $u_{1,f}/v_1=4.5, u_{2,f}/v_1=-1$ for (c). (d)Topological phase transition time, prediected by $t_{p,0}$, varies with the post-quench parameter $u_{1,f}$.
}
\label{fig:fig2}
\end{figure}

\begin{figure}[t]
\includegraphics[width=8cm]{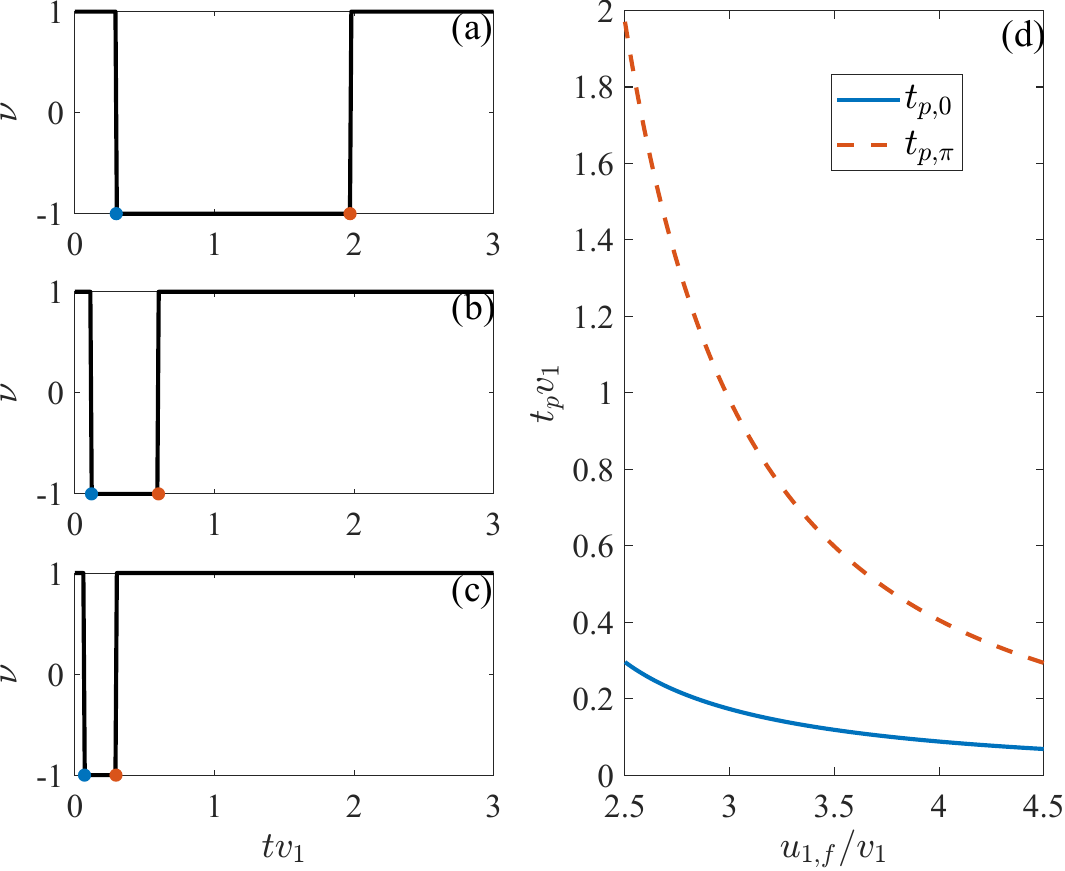}
\caption{(a)(b)(c)Topological phase transition characterized by the modular Hamiltonian's topological invariant $\nu(t)$ following a dissipative quench from Lindbladian steady state of  $(M_0,M_\pi)=(-1,-1)$ topological phase to those in $(M_0,M_\pi)=(1,1)$ phase. During the entire quench process, the ratio $v_2/v_1$ is maintained at $-2$, and the dimensionless parameter $\gamma$ takes the value $0.1$. Initial density matrix is the steady state of the Lindbladian with parameters $u_{1,i}/v_1=1$, $u_{2,i}/v_1=-1$ and the post-quench Lindbladian parameters are taken as $u_{1,f}/v_1=2.5, u_{2,f}/v_1=-1$ for (a), $u_{1,f}/v_1=3.5, u_{2,f}/v_1=-1$ for (b), $u_{1,f}/v_1=4.5, u_{2,f}/v_1=-1$ for (c). (d)Critical time of topological phase transition prediected by $t_{p,0}$ and $t_{p,\pi}$ vary with the post-quench parameter $u_{1,f}$ .
}
\label{fig:fig3}
\end{figure}

These theoretical predictions are in agreement with the numerical simulations shown in Figs.~\ref{fig:fig2}-\ref{fig:fig3}. The quench from $(M_0,M_\pi)=(1,-1)$ to $(1,1)$ (Fig.~\ref{fig:fig2}) produces a single transition at $t_{p,0}$. The quench from $(M_0,M_\pi)=(1,-1)$ to $(-1,1)$ (Fig.~\ref{fig:fig3}) generates two distinct transitions at $t_{p,0}$ and $t_{p,\pi}$. For both cases, the transition time is precisely predicted by Eq.~\eqref{tpif}. The remarkable consistency between analytical predictions and numerical results provides strong validation of our theoretical framework.


\section{Entanglement spectrum dynamics}
\label{sec:Entanglement}
\begin{figure}[t]
\includegraphics[width=8cm]{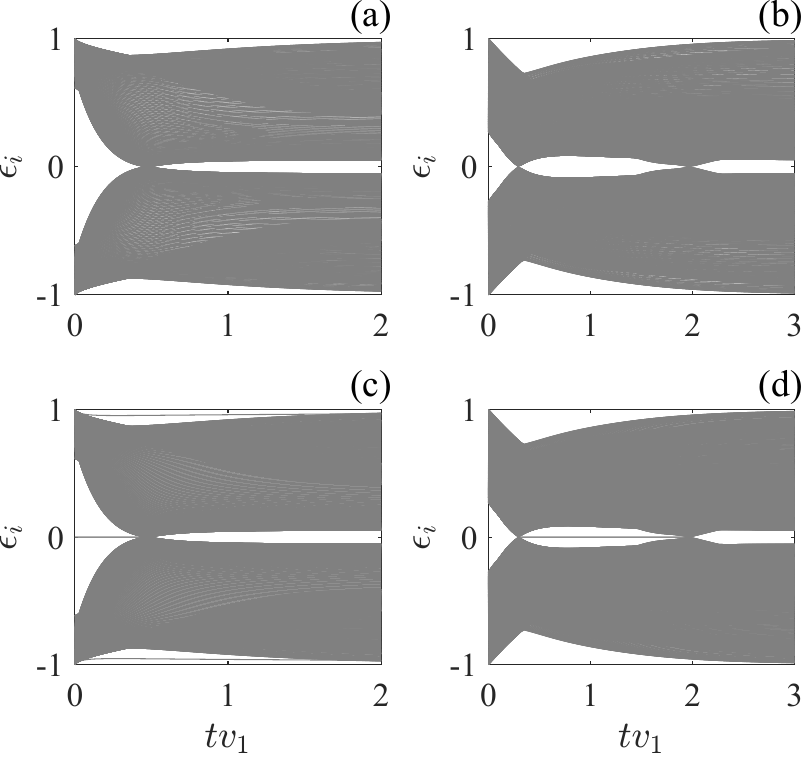}
\caption{Time-evolution of single particle entanglement spectrum. (a)(b) shows the single particle entanglement spectrum under periodical boundary conditions with gap closing dynamics and (c)(d) shows the single particle entanglement spectrum under open boundary conditions (OBC) with the topologically protected zero-mode structure. In (a)(c), the parameters is taken the same as Fig.\ref{fig:fig2}(a) and the time-evolving density matrix undergoes one topological phase transitions with its topological invariant changing from $-1$  to $+1$. In (b)(d), the parameters is taken the same as Fig.\ref{fig:fig3}(a) and the time-evolving density matrix undergoes two topological phase transitions with its topological invariant changing from $1$ to $-1$, and sequentially from $-1$ to $1$. Also, we have taken the system Hamiltonian of Eq.\eqref{eqlind} as $\hat{H}=J\hat{c}_m^\dagger\hat{c}_{m+1}+J\hat{c}_{m+1}^\dagger\hat{c}_{m}$ with $J/v_1=1$ and the system size $N=500$ in this figure.
}
\label{fig:fig4}
\end{figure}
%{\color{red}Citations Here Need Fixing}
The characterization of topology through many-body entanglement spectrum analysis has been well-established in closed quantum systems \cite{ESQuench2,ESQuench3,ESQuench4}. Recent extensions to open quantum systems have revealed intriguing connections between entanglement spectrum crossings and non-Hermitian topology in Lindbladian dynamics\cite{ESLind2}. Here, we establish a direct correspondence between the topology of mixed states, encoded in their modular Hamiltonians, and the entanglement spectrum structure. %(Specifically, our dynamical studies demonstrate that topological phase transitions in time evolution coincide with gap closures under periodic boundary conditions and zero-mode eigenvalue emergence under open boundary conditions in single-particle entanglement spectrum. )

For a Gaussian state, the entanglement spectrum, defined by the eigen values of density matrix $\hat{\rho}$, is fully determined by the correlation matrix $\Delta(t)$. The computation proceeds by first diagonalizing $\Delta(t)$ to obtain its eigenvalues, known as the single-particle entanglement spectrum (SPES) ${\epsilon_i}$, which results in a set of real eigenvalues due to the matrix's Hermiticity. Subsequently, the full many-body entanglement spectrum $\Xi_{{\mathbf{e}}}$ is built from these single-particle eigenvalues via the relation 
\begin{equation}
\Xi_{\{\mathbf{e}\}}=\prod_{i=1}^{N } \frac{1}{2}\left[1+(-1)^{e_i} \epsilon_i\right],
\end{equation}
where ${\mathbf{e}} = (e_1, e_2, \dots, e_N)$ with $e_i \in \{0,1\}$ denotes the occupation configuration of the single-particle modes\cite{ESQuench3}. In our numerical simulations, we compute $\Delta(t)$ at each time step by solving Eq.\eqref{delt} and subsequently diagonalize it to obtain the SPES. This procedure ensures that the topological features of the state are directly accessible from the SPES.

We present numerical simulations of the modular Hamiltonian's spectral evolution in Fig.~\ref{fig:fig4}. The PBC spectra, shown in Fig.~\ref{fig:fig4}(a)(b), exhibit characteristic gap closure at critical times $t_{p,0/\pi}$, revealing direct correspondence between single particle entanglement spectra and topological phase transitions. The OBC spectra, shown in Fig.~\ref{fig:fig4}(c)(d), manifests robust zero-mode eigenvalues when the time-dependent density matrix is topologically non-trivil and the $Z_2$ topological invariant of modular Hamiltonian is $-1$. These results establish a concrete manifestation of the bulk-boundary correspondence principle for nonequilibrium density matrices through their entanglement spectral signatures.

\section{Summary}
\label{sec:sum}
We have studied a class of quadratic Lindblad dynamics for understanding dissipation-driven topological phase transitions in a one-dimensional quantum open system. These steady states belong to class D. Their topology, classified by a $Z_2$ Pfaffian invariant, is shown to depend exclusively on the dissipation operators, with no dependence on the system Hamiltonian. This universality arises from the symmetry-constrained structure of the modular Hamiltonian at high-symmetry momenta. Under quench dynamics, multiple topological transitions emerge at precisely predictable critical times, governed solely by the dissipation parameters, even when initial and final steady states share identical topological indices. The entanglement spectrum evolution further reveals a robust bulk-edge correspondence: gap closures in the single-particle spectrum under periodic boundary conditions coincide with topologically protected zero modes in open boundaries. These results unambiguously link the non-equilibrium topology of density matrices to experimentally measurable entanglement signatures, offering a pathway to dynamically probe and control dissipation-induced topology in quantum simulation platforms. Our findings highlight dissipation as a dominant resource for engineering topological phases beyond equilibrium constraints.
\appendix
\section{Derivation of Eq.~\eqref{Delsksum}}
\label{app:Delsksum}

We firstly employ a vectorization procedure\cite{vect1}for the steady-state equation for the correlation matrix Eq.~\eqref{delsk}. The vectorization map $\text{vec}(\cdot)$ transforms a matrix into a column vector by stacking its columns. A fundamental property of this map is:
\begin{equation}
    \text{vec}(A \rho B) = (A \otimes B^T) \text{vec}(\rho),
\end{equation}
which, in the Dirac notation used in our manuscript ($|\rho\rangle \equiv \text{vec}(\rho)$), becomes:
\begin{equation}
A \rho B \rightarrow (A \otimes B^T) |\rho\rangle.
\end{equation}

Applying this property term by term to our steady-state equation:
\begin{align}
& H_{\text{eff}} \Delta_s \rightarrow (H_{\text{eff}} \otimes I) |\Delta_s\rangle, \nonumber\\
& \Delta_s H_{\text{eff}}^{\dagger} \rightarrow (I \otimes (H_{\text{eff}}^{\dagger})^T) |\Delta_s\rangle = (I \otimes H_{\text{eff}}^*) |\Delta_s\rangle, \nonumber\\
& 4i Y \rightarrow 4i |Y\rangle,
\end{align}
then Eq.~\eqref{delsk} is transformed into 
\begin{equation}  
\label{Delvec}
|\Delta_s\rangle = 4i \left[ H_{\text{eff}} \otimes I - I \otimes H_{\text{eff}}^* \right]^{-1} |Y\rangle.  
\end{equation}

The non-Hermitian Hamiltonian $H_{\rm{eff}}$ is decomposed as 
\begin{equation} 
\label{Heffdec}
H_{{\rm eff}}=\lambda_{+}|\psi_{+}\rangle\langle\chi_{+}|+\lambda_{-}|\psi_{-}\rangle\langle\chi_{-}|,
\end{equation}
where $|\psi_\pm\rangle$ and $|\chi_\pm\rangle$ are the right and left eigen-vectors defined by $H_{\rm{eff}}|\psi_\pm\rangle=\lambda_\pm|\psi_\pm\rangle$ and $H^\dagger_{\rm{eff}}|\chi_\pm\rangle=\lambda^*_\pm|\chi_\pm\rangle$.  Substituting Eq.~\eqref{Heffdec} into Eq.~\eqref{Delvec}, $|\Delta_s\rangle$ is derived as
\begin{align} 
\label{Delsvec2}
&|\Delta_{s}\rangle=i4[(\sum_{\alpha=\pm}\lambda_{\alpha}|\psi_{\alpha}\rangle\langle\chi_{\alpha}|)\otimes(\sum_{\beta=\pm}|\psi_{\beta}^{*}\rangle\langle\chi_{\beta}^{*}|)\nonumber\\
&-(\sum_{\alpha=\pm}|\psi_{\alpha}\rangle\langle\chi_{\alpha}|)\otimes(\sum_{n=\pm}\lambda_{\beta}^{*}|\psi_{\beta}^{*}\rangle\langle\chi_{\beta}^{*}|)]^{-1}|Y\rangle\nonumber\\
&=\sum_{\alpha,\beta=\pm}\frac{1}{(\lambda_{\alpha}-\lambda_{\beta}^{*})}|\psi_{\alpha}\rangle\langle\chi_{\alpha}|\otimes|\psi_{\beta}^{*}\rangle\langle\chi_{\beta}^{*}|]|Y\rangle,
\end{align}
where we have used the completeness relation $\sum_{\alpha=\pm}|\psi_{\alpha}\rangle\langle\chi_{\alpha}|$. Rewriting Eq.~\eqref{Delsvec2} in the form of matrix, Eq.~\eqref{Delsksum} is obtained.

\section{Proof of Eq.~\eqref{delt0pff}}
\label{app:delt0pff}
To rigorously establish Eq.~\eqref{delt0pff}, we first write down the eigenvalues of $H_{\rm{eff}}(k_s)$ as
\begin{align}  
&\lambda_\pm(k_s) = ih_0(k_s) \pm E(k_s), \nonumber\\
&E(k_s) = \sqrt{h^2_y(k_s) - h^2_x(k_s) - h^2_z(k_s)}.  
\end{align}  
Considering an eigenstate $|\psi_+\rangle$, we compute 
\begin{equation} 
\label{Hkspsi}
H_{\rm{eff}}(k_s)|\psi_+\rangle =\lambda_+|\psi_+\rangle.  
\end{equation}  
The complex conjugate of Eq.\eqref{Hkspsi} gives
\begin{equation}  
H^*_{\rm{eff}}(k_s)|\psi_+^*\rangle =\lambda_+^*(k_s)|\psi_+^*\rangle. 
\end{equation}  
Under the constraint $H^*_{\rm{eff}}(k_s) = -H_{\rm{eff}}(k_s)$, it becomes  
\begin{equation}  
H_{\rm{eff}}(k_s)|\psi_+^*\rangle =-\lambda^*_+(k_s)|\psi_+^*\rangle = [ih_0(k_s)-E^*(k_s)]|\psi_+^*\rangle.  
\end{equation}  
This indicates $|\psi_+^*\rangle$ is the eigen-vector of $H_{\rm{eff}}$ with eigenvalue $ih_0(k_s) - E^*(k_s)$.
When $h^2_y(k_s)>h^2_x(k_s)+h^2_z(k_s)$, $E(k_s)$ is purely real, where $H_{\rm{eff}}$ lies in the parity-time (PT)-symmetry preserved regime. In this regime, $|\psi_+^*\rangle$ is the eigenstate of $H_{\rm{eff}}(k_s)$
with eigenvalue $\lambda_-$, which means $|\psi_+^*\rangle=|\psi_-\rangle$. For the PT-symmetry-broken regime $h^2_y(k_s)<h^2_x(k_s)+h^2_z(k_s)$, $|\psi_+^*\rangle$ is the eigenstate of $H_{\rm{eff}}$ with eigenvalue $\lambda_+$, which means $|\psi_+^*\rangle=|\psi_+\rangle$.

For the PT-symmetry-preserved regime, explicitly writing $|\psi_+\rangle = (a, b)^T$, we obtain $|\psi_-\rangle = (a^*, b^*)^T$, which leads to 
\begin{equation}  
\langle \psi_- | \sigma_y | \psi_+ \rangle = \left(\begin{array}{cc}  
a & b  
\end{array}\right) \sigma_y \left(\begin{array}{c}  
a \\  
b  
\end{array}\right) = 0.  
\end{equation}  
By analogous reasoning, we have
\begin{align}  
\label{fmnPTp}
\langle \psi_+ | \sigma_y | \psi_- \rangle &= 0,\nonumber\\
\langle \chi_+ | \sigma_y | \chi_- \rangle &= 0, \nonumber\\  
\langle \chi_- | \sigma_y | \chi_+ \rangle &= 0.  
\end{align}  
Consequently, the overlap factors satisfy  
\begin{align}  
\label{fpmpre}
f_{+,-} &= 0, \quad f_{-,+} = 0, \nonumber\\  
f_{+,+} &= 1, \quad f_{-,-} = 1.  
\end{align}  

For the PT-symmetry-broken regime, we employ a derivation analogous to that of PT-symmetry-preserved case. It is derived to $\langle \psi_+ | \sigma_y | \psi_+ \rangle = 0$ from $|\psi_+^*\rangle = |\psi_+\rangle$, which leads to
\begin{align}  
\label{fmnPTb}
\langle \psi_- | \sigma_y | \psi_- \rangle &= 0,\nonumber\\
\langle \chi_+ | \sigma_y | \chi_+ \rangle &= 0, \nonumber\\  
\langle \chi_- | \sigma_y | \chi_- \rangle &= 0,
\end{align}  
and
\begin{align}  
\label{fpmbro}
f_{+,+} &= f_{-,-} = 0, \nonumber\\  
f_{+,-} &= f_{-,+} = 1.  
\end{align}  
We substitute the respective overlap factors $f_{\pm,\pm}$ of the PT-symmetric regime and the PT-broken regime into the the Pfaffian's expression Eq. \eqref{Mkchipsi}. Since $\lambda_+(k_s)-\lambda_{\pm}^*(k_s)=2ih_0$ holds for $E(k_s)=\pm E^*(k_s)$, this substitution process yields identical expressions for the Pfaffian of the correlation matrix in both regimes, with the final unified result being explicitly given in Eq.~\eqref{delt0pff}.

{\bf Acknowledgements}
This work was supported by the National Natural Science Foundation of China
(grant NO. 12204406 and Grand No. 12504187), 
the Yunnan Fundamental Research Projects (grant NO. 202401CF070187),
and the Start-up Research Fund of Southeast University (RF1028624190).

{\bf Code availability.} Codes for numerical calculations and figures are available at \cite{code}.

\bibliography{ref.bib}

\end{document}